\documentclass[prl,superscriptaddress,amsmath,amssymb,floatfix,margin=1in,super,twocolumn]{revtex4-1} 

\usepackage[dvips]{graphicx}
\usepackage{xspace}
\usepackage[usenames,dvipsnames]{xcolor}
\usepackage{array}
\usepackage{ulem}
\usepackage{hyperref}
\usepackage[english]{babel}
\usepackage{csquotes}

\setcitestyle{super}

\def\app#1#2{%
  \mathrel{%
    \setbox0=\hbox{$#1\sim$}%
    \setbox2=\hbox{%
      \rlap{\hbox{$#1\propto$}}%
      \lower1.1\ht0\box0%
    }%
    \raise0.25\ht2\box2%
  }%
}
\def\approxprop{\mathpalette\app\relax}

\newcommand{\COSO}{Cu$_{2}$OSeO$_{3}$\@\xspace}

\newcommand{\phii}{Institute of Physics 2, Faculty of
Mathematics and Natural Sciences, University of Cologne, Z\"{u}lpicher Stra{\ss}e 77, D-50937 Cologne, Germany}
\newcommand{\mineralogy}{Institute of Geology and Mineralogy, Faculty of
Mathematics and Natural Sciences, University of Cologne, Z\"{u}lpicher Stra{\ss}e 49b, D-50674 Cologne, Germany}

\begin{document}

\title{Coupled dynamics of long-range and internal spin cluster order in Cu$_{2}$OSeO$_{3}$}

\author{Rolf~B.~Versteeg}
 \email[Corresponding author:]{versteeg@ph2.uni-koeln.de}
\author{Jingyi~Zhu}
\author{Christoph~Boguschewski}
\author{Fumiya~Sekiguchi}
\author{Anuja~Sahasrabudhe}
\author{Kestutis~Budzinauskas}
\author{Prashant~Padmanabhan}  
  \affiliation{\phii}
\author{Petra~Becker}  
\affiliation{\mineralogy}
\author{Daniel~I.~Khomskii}
\affiliation{\phii}
\author{Paul~H.~M.~van Loosdrecht} 
 \email[Corresponding author:]{pvl@ph2.uni-koeln.de}
\affiliation{\phii}

\date{\today}

\begin{abstract}
Cu$_4$ triplet clusters form the relevant spin entity for the formation of long-range magnetic order in the cluster magnet Cu$_2$OSeO$_3$. Using time-resolved Raman spectroscopy, we probed photoinduced spin and lattice dynamics in this Mott insulator. Multiple ps-decade spin-lattice relaxation dynamics is observed, evidencing a separation of the order parameter dynamics into disordering of long-range and internal spin cluster order.
Our study exemplifies the double order parameter dynamics of generalized molecular crystals of charge, spin, and orbital nature. 
\end{abstract}

\maketitle

Quantum materials with at least two length scales for electronic interactions lead to the self-formation of generalized molecular crystals of charge, spin, and orbital nature.\cite{attfield2015orbital,streltsov2017orbital} From strong, shorter length scale electronic interactions solid-state \enquote{molecules} or clusters form, which \enquote{crystallize} through weaker, longer length scale electronic interactions. Such emergent solid-state molecular ground states have been identified in, for instance, the mineral Fe$_3$O$_4$ where trimeron orbital molecules form, \cite{senn2012charge,dejong2013speed} molybdenates with triangular orbital plaquette molecules,\cite{sheckelton2012possible,mccarroll1957some} and the chiral magnet \COSO where tetrahedral spin triplets form.\cite{clustercoso} The order-to-disorder phase transition pathways in such quantum materials comprises disordering of both the order inside the individual cluster actors, as well as the emerging long-range cluster order. 
An understanding of such dynamic double order parameter behaviour
is important for the fundamental understanding of electronic correlations in solid-state molecular ground states and from the perspective of potential switching applications based on spin, charge and orbital degrees of freedom.\cite{dejong2013speed,wall2018ultrafast,kirilyuk2010}

In this context we investigate the nonequilibrium dynamics of spin cluster and long-range order in the cluster Mott insulator \COSO. Nonequilibrium dynamic studies of \COSO have predominantly focused on the chiral magnetism, in particular the manipulation and excitation of the metamagnetic Skyrmion phase.\cite{seki2012,ogawa2015,langner2017,white2018electric} An equally intriguing aspect of the nonequilibrium dynamics concerns the disordering pathways of long-range and spin triplet cluster order. The nonequilibrium dynamics is governed by lattice and spin excitations of long-range and internal cluster
character and their coupling. This is expected to lead to rather rich and interesting dynamic behaviour, which can be mapped in the time-domain by ultrafast spectroscopy techniques.

\begin{figure}[h!]
\center
\includegraphics[width=3.375in]{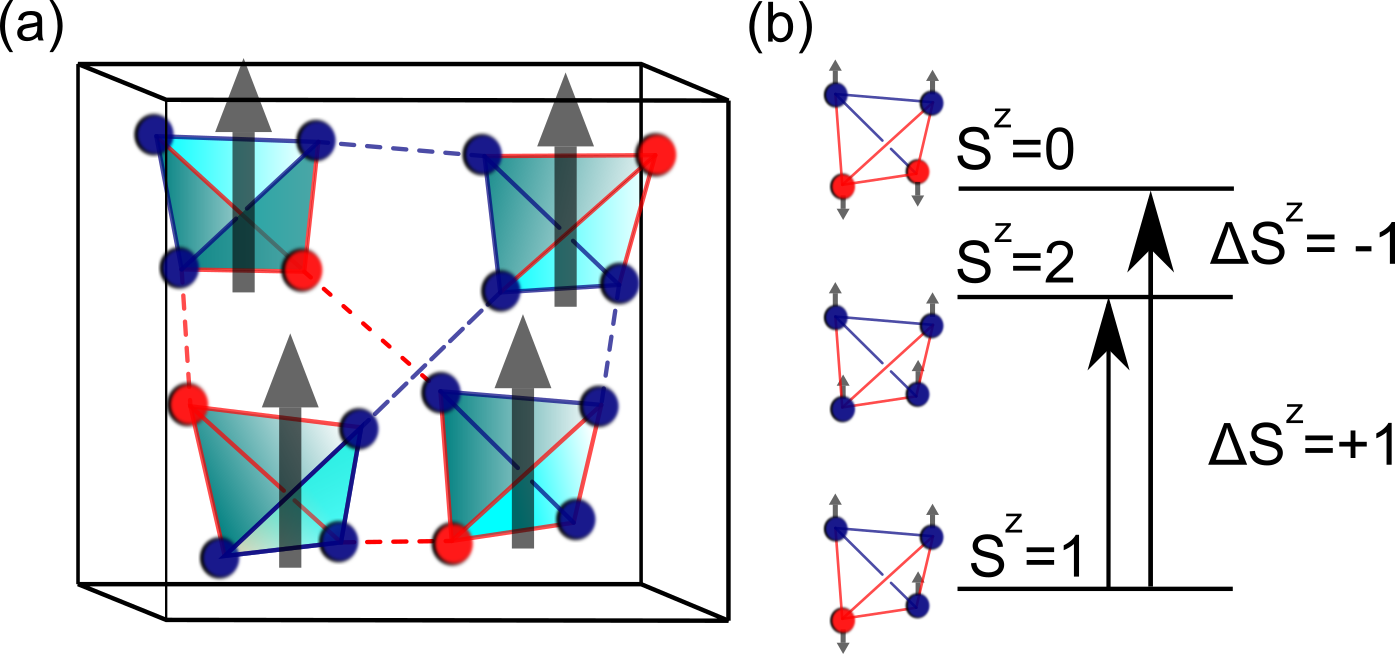}
\caption{
The magnetic unit cell of \COSO. The red (spin-down) and blue (spin-up) Cu$^{2+}$ $S$\,$=$\,$\frac{1}{2}$ spins reside on a distorted pyrochlore lattice. The exchange paths can be classified into strong (full lines) and weak (dashed line) exchange couplings. The blue lines represent a predominantly ferromagnetic exchange, whereas the red lines correspond to a predominantly antiferromagnetic exchange. The three-up-one-down spin triplet ($S$\,$=$\,$1$) clusters are indicated in light blue, with spin in grey. 
b) Spin flip excitations of a triplet tetrahedron. The spin flip leads to a raise or lowering of the cluster's total magnetic quantum number $S^z$.
}
\label{fig:together}
\end{figure}

The spin cluster formation in \COSO results from the geometric distortion away from a perfect magnetic pyrochlore lattice.\cite{clustercoso,hopkinson2006} The magnetic unit cell is shown in Fig.\,\ref{fig:together}a, and consists of 16 Cu$^{2+}$  $S$\,=\,$\frac{1}{2}$ spins on a distorted pyrochlore lattice.\cite{jwbos2008} Long and short Cu$^{2+}$-Cu$^{2+}$ path ways are identified with correspondingly weak (dashed lines) and strong (full lines) exchange interactions. The $S$\,=\,$1$ spin clusters form through the strong intra-cluster exchange interactions far above the long-range ordering temperature $T_{C}$\,$\approx$\,$58$\,K, below which the weak inter-cluster exchange interactions order the Cu$_4$ spin clusters into a spin cluster helix with a chiral pitch of approximately $62$\,nm.\cite{clustercoso,tucker2016,longwavelength} The cluster magnetic order leads to distinctly different types of low- and high-energy spin excitations with long-range and internal Cu$_4$ cluster character. The high-energy cluster modes can be understood as spin flip excitations of the triplet tetrahedron, as illustrated in Fig.\;\ref{fig:together}b. The collective motion of the ordered triplet clusters gives rise to spin excitations at low energy, including the Goldstone mode of the magnetically long-range ordered state. 
\cite{romhanyi2014,ozerov2014,portnichenko2016}

Time-resolved spontaneous Raman spectroscopy \cite{versteeg2018,faustibook} allowed us to synchronously probe photoinduced spin and lattice dynamics in \COSO. Our results reveal an efficient coupling of the high-energy spin cluster excitations to optical phonons, and more importantly, that the photoinduced magnetization dynamics is governed by a double order parameter behaviour reflecting both the disordering of long-range and internal spin cluster order.

Before turning to the the non-equilibrium optical spectroscopy results, we discuss the equilibrium inelastic light scattering response of \COSO. Single crystals of a few mm$^3$ size were synthesized by chemical transport reaction growth.\cite{belesi2010ferrimagnetism} 
A (111) oriented plate-shaped sample with a flat as-grown face was used. The Raman probing is realized with $2.42$\,eV pulses of $\Delta\nu$\,$\approx$\,$1.2$\,meV ($\approx$\,$10$\,cm$^{-1}$) bandwidth (full-width at half-maximum, FWHM) and $\Delta\tau$\,$\approx$\,$1.5$\,ps (FWHM) duration. The probe beam polarization is parallel to the crystallographic [1$\bar{1}$0] axis. The probing is carried out in an unpolarized Raman geometry in order to optimize the scattered light detection efficiency. The energy resolution typically lies around $10$\,cm$^{-1}$ and is largely dictated by the probe pulse bandwidth.

\begin{figure}[h]
\center
\includegraphics[width=3.375in]{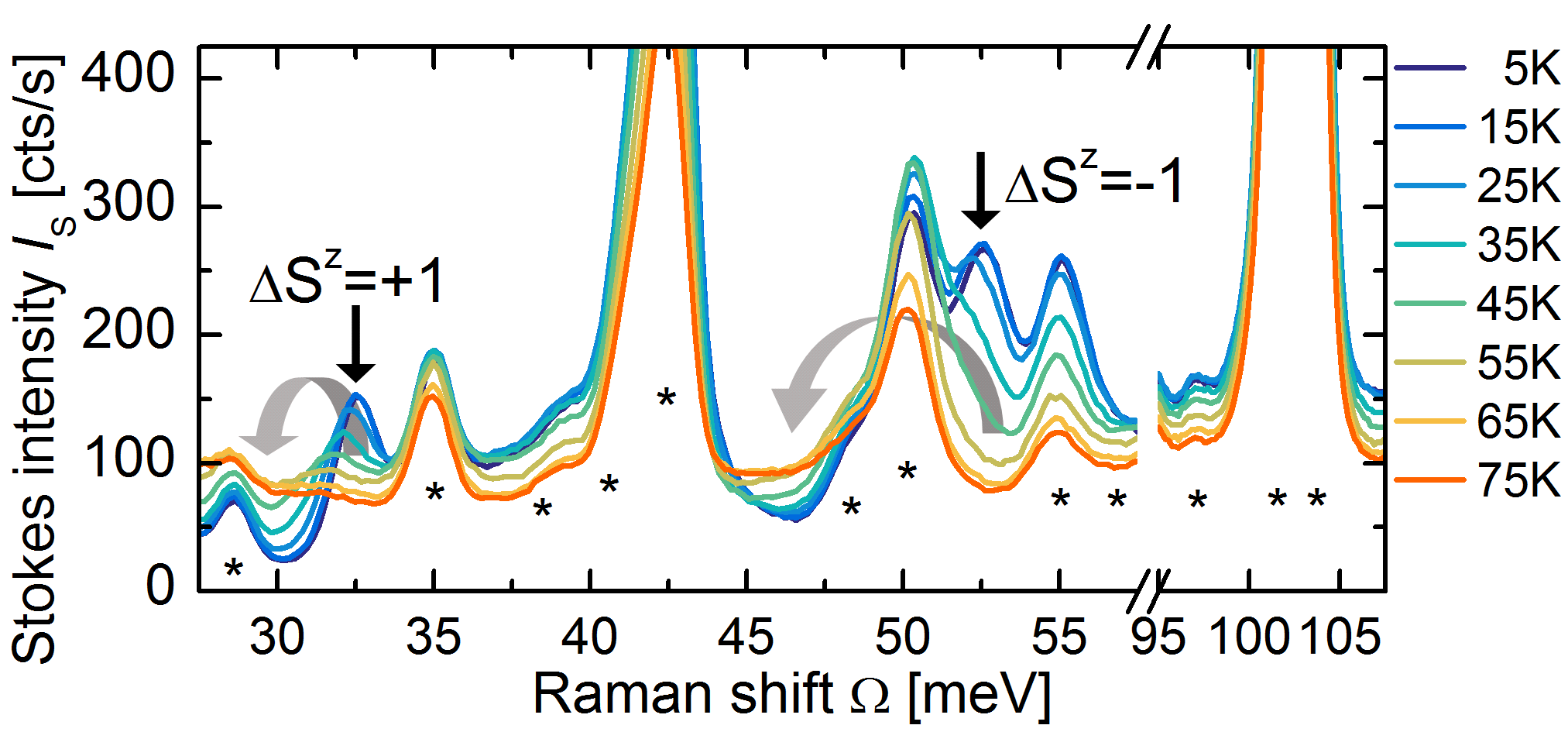}
\caption{Temperature-dependent steady-state Raman spectra. The modes indicated with an asterisk (*) are phonons, or not fully resolved phonon regions. 
High-energy spin cluster excitations are observed at $\mathit{\hbar\Omega}$\,$\sim$\,$32$\,meV ($\Delta S^z$\,=\,$+$\,$1$) and $\mathit{\hbar\Omega}$\,$\sim$\,$53$\,meV ($\Delta S^z$\,=\,$-$\,$1$). 
}
\label{fig:steady}
\end{figure}

Figure \ref{fig:steady} shows the steady-state Raman spectra at temperatures ranging from $5$\,K to $75$\,K. The large amount of atoms in the structural unit cell results in a rich set of phonon modes, as indicated with an asterisk in the figure. \cite{gnezdilov2010,miller2010,kurnosov2012} Two strongly temperature dependent modes are observed, which were previously assigned to $\Delta S^z$\,=\,$+$\,$1$ ($\mathit{\hbar\Omega}$\,$\sim$\,$32$\,meV) and $\Delta S^z$\,=\,$-$\,$1$ ($\mathit{\hbar\Omega}$\,$\sim$\,$53$\,meV) high-energy spin cluster excitations at the center of the Brillouin zone. \cite{romhanyi2014,ozerov2014,portnichenko2016} For these modes, which are Raman-active through the Elliot-Loudon scattering mechanism, \cite{cottam1986light,fleuryloudon1968} a spectral weight transfer to lower Stokes shift $\mathit{\Omega}$ is observed when the temperature increases towards $T_{C}$, as indicated with the grey arrows. The magnetic spectral weight transfer is understood as a softening and broadening of the $\Delta S^z$\,=\,$\pm$\,$1$ spin excitations. Above $T_{C}$\,$\approx$\,$58$\,K magnetic scattering still persists, however as a broad scattering continuum. \cite{kurnosov2012} Similar critical behaviour was observed for a spin cluster transition in the THz-range of the absorption spectrum. \cite{laurita2017}

The conceptual picture of spin cluster excitations indicated in Fig.\,\ref{fig:together}b applies well above $T_{C}$ where no long-range order exists between the clusters. In this limit the high-energy spin excitations can be regarded as dispersionless magnetic excitons with a resultingly broad continuum light scattering spectrum. \cite{liu1976magnetic,birgenau1971} In the long-range ordered phase these internal cluster modes acquire dispersion by the inter-cluster exchange interactions, and form optical magnon branches, resulting in well-defined magnetic modes in the Raman spectrum.\cite{samuelsen1974,portnichenko2016} The Raman-active high-energy $\Delta S^z$\,=\,$\pm$\,$1$ spin excitations thus form an optical probe for both the long-range and internal spin cluster order.

We now turn to the optically induced spin and lattice dynamics.
The \COSO sample, cooled to $5$\,K, is excited 
in the crystal-field excitation region\cite{versteeg2016}
with $2.18$\,eV pump pulses of $\Delta\tau$\,$\approx$\,$0.3$\,ps (FWHM) duration at a fluence of F\,$\approx$\,$2$\,mJ/cm$^{2}$. The fraction of photoexcited Cu$^{2+}$-sites per pulse lies around $\sim$\,$5\cdot 10^{-6}$. The weak pump-excitation conditions ensure that we probe the near-equilibrium dynamics of the helimagnetic phase. The thermalization of the system is measured by the time-evolution of the Stokes spectrum. 
The Raman-probe pulse falls in the low energy tail of the charge transfer excitation region. \cite{versteeg2016} The Stokes Raman scattering intensity $I_{\rm S}$($\Omega$) is proportional to $I_{\rm S}(\mathit{\Omega})$\,$\propto$\,$V_{\rm probe}[\alpha(\omega)]\cdot \chi^2_{\rm R}(\mathit{\Omega})\cdot [n(\mathit{\Omega})+1]$ (Ref.\,\citenum{compaan1984}). Here $V_{\rm probe}[\alpha(\omega)]$ describes the probe volume term, which depends on the absorption coefficient $\alpha(\mathit{\omega})$ at the scattered photon frequency $\mathit{\omega}$. $\chi^2_{\rm R}(\mathit{\Omega})$ gives the squared Raman tensor and $[n(\mathit{\Omega})+1]$ the population factor. Under the weak pump excitation conditions the (transient) occupation number $n(\mathit{\Omega})$\,$\ll$\,$1$ and can thus be neglected, i.\,e. $I_{\rm S}(\mathit{\Omega})$\,$\approxprop$\,$V_{\rm probe}[\alpha(\mathit{\omega})]\cdot \chi^2_{\rm R}(\mathit{\Omega})$. 

\begin{figure*}[t]
\center
\includegraphics[scale=1]{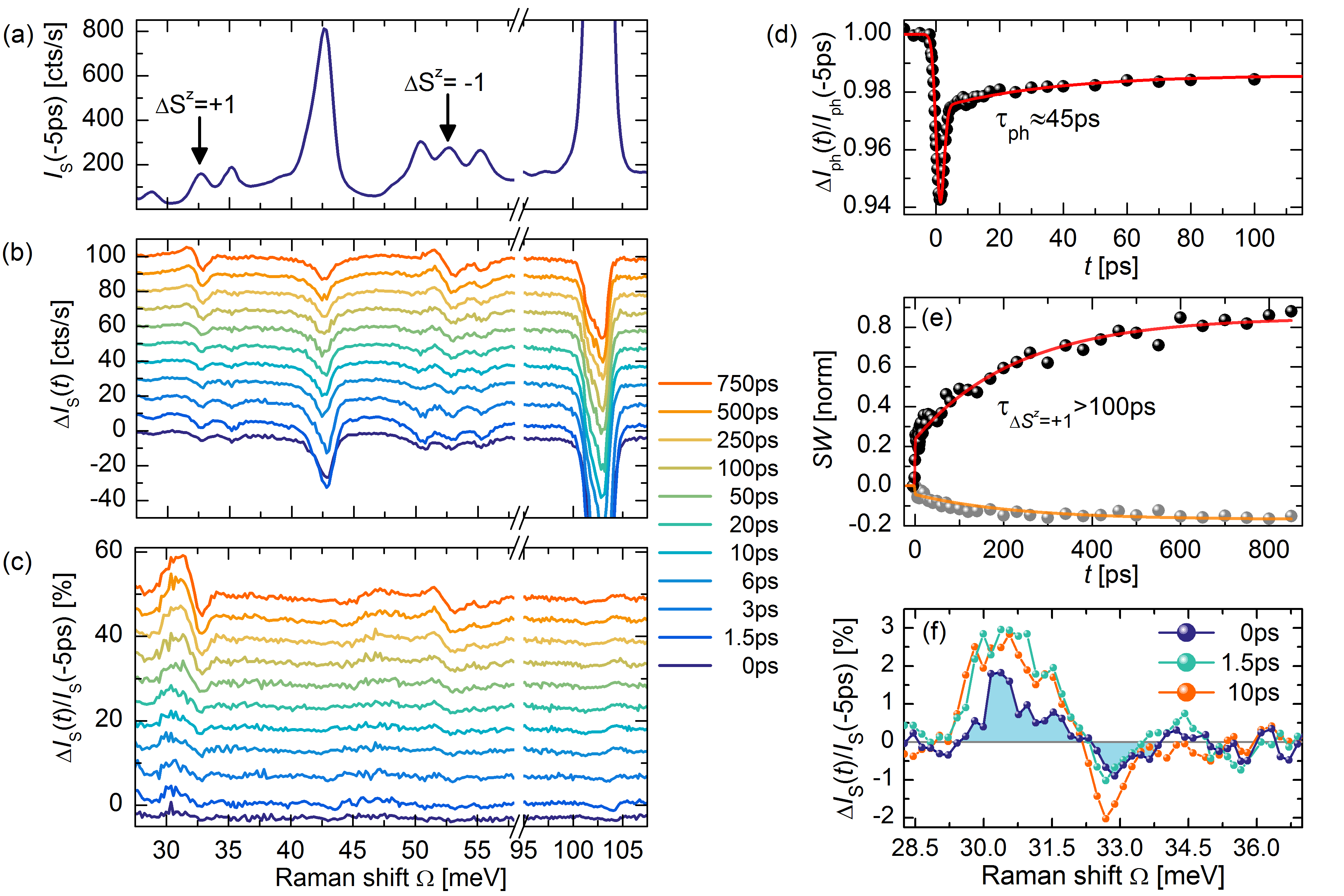}
\caption{a) Stokes spectrum at $5$\,K. The spin cluster excitations are indicated with arrows. The other peaks are phonons. b) Differential Stokes spectra for indicated time-delays. The phonon modes show a reduction in scattering efficiency, which does not fully recover within the measured time window. The $\Delta S^z$\,=\,$\pm$\,$1$ spin excitations show a dynamic softening and broadening. c) Scaled differential spectra. From these spectra it becomes clear that the phonons only show a negligible frequency shift, in sharp contrast with the $\Delta S^z$\,=\,$\pm$\,$1$ spin excitations. After $t$\,$>$\,$100$\,ps the spin excitation spectral weight transfer keeps on increasing. d) $\Delta I_{\rm ph}(t)$/$I_{\rm ph}$($-5$\,ps) transient for the $103$\,meV phonon region. An initial scattering reduction and partial recovery within the time-resolution is observed, in addition to a longer time-scale recovery. Global fitting approximately gives the same long recovery time-constant of $\tau_{\rm ph}\sim$\,$45$\,ps for all phonons. e) The softening of the $\Delta S^z$\,=\,$+$\,$1$ spin excitation occurs on a significantly slower timescale of $\tau_{\Delta S^z=\pm 1}$\,$>$\,$100$\,ps. Plotted is the increasing (black dots) and decreasing (grey dots) component of the spectral weight transfer. f) Differential Stokes spectra $\Delta I_{\rm S}(\Omega,t)$/$I_{\rm S}$($\mathit{\Omega}$,$-5$\,ps) of the $\Delta S^z$\,=\,$+$\,$1$ excitation at early delay times. At pump-probe overlap ($t$\,$=$\,$0$\,ps) a spectral weight shift is established, revealing an efficient energy relaxation channel on the shortest time-scale ($<$\,$1.5$\,ps).}
\label{fig:panel}
\end{figure*}

In Fig.\,\ref{fig:panel} we show our main result: the transient evolution of the Stokes spectrum $I_{\rm S}$($\mathit{\Omega}$,$t$), where $t$ refers to the pulse delay time and $\mathit{\Omega}$ to the Stokes shift. For clarity  the  pre-time-zero Stokes spectrum $I_{\rm S}$($\mathit{\Omega}$,$-5$\,ps) is plotted in Fig.\,\ref{fig:panel}a.
Figures \ref{fig:panel}b and \ref{fig:panel}c show the differential Stokes spectra $\Delta I_{\rm S}(\mathit{\Omega},t)$\,=$I_{\rm S}(\mathit{\Omega},t)$\,$-$\,$I_{\rm S}$($\mathit{\Omega}$,$-5$\,ps),
and scaled differential Stokes spectra $\Delta I_{\rm S}(\Omega,t)$/$I_{\rm S}$($\mathit{\Omega}$,$-5$\,ps) for the time-delays indicated in the figure. The two main observations are a decrease with subsequent recovery of the phonon-scattering intensity, and a transient broadening and spectral weight transfer to lower $\mathit{\Omega}$ of the 
$\Delta S^z$\,=\,$\pm$\,$1$ spin excitations. Similar behaviour is observed at higher bias temperatures.

The scattering intensity of selected phonon (ph) modes is integrated over a range $\sim$\,$3\times$FWHM centered at the phonon energy to give $I_{\rm ph}(t)$, and shown in Fig.\,\ref{fig:panel}d as relative transient phonon scattering intensity $\Delta I_{\rm ph}(t)$/$I_{\rm ph}$($-5$\,ps). For all phonons an initial Stokes scattering efficiency decrease of $\sim$\,$5$\% after excitation is observed, with a partial recovery within the temporal pump-probe pulse overlap. A slower recovery time-scale of $\tau_{\rm ph}$\,$\sim$\,$45$\,ps is observed to $\Delta I_{\rm ph}(t)$/$I_{\rm ph}$($-5$\,ps)\,$\approx$\,$-1.5$\% at late delay times. The phonon spectral shape hardly changes, as most clearly seen in the scaled differential Stokes spectra (Fig.\,\ref{fig:panel}c). From these observations it becomes apparent that the Raman spectra show an overall reduction and recovery in scattering efficiency due to a transient change in absorption, with a concomitant change in $V_{\rm probe}[\alpha(\mathit{\omega})]$.

A transient softening and broadening of the $\Delta S^z$\,=\,$\pm$\,$1$ spin excitation scattering is observed due to excitation of the magnetic system. Note that the asymmetric line shape in the (scaled) differential spectra originates from a line shape change in $\chi^2_{\rm R}$, and not from the population term $n(\mathit{\Omega})$.
The time dependent dynamics of the spin excitation scattering is convoluted with the change
in the probing volume.
This effect is deconvoluted by scaling the spectrum with the scattering intensity of the  
phonon response, giving the intensity ratio:

\begin{equation}
\frac{I_{\Delta S^z=\pm 1}}{I_{\rm ph}}\propto\frac{\chi^2_{\Delta S^z=\pm 1}}{\chi^2_{\rm ph}}
\label{eq:stokesratio}
\end{equation}

The intensity ratio is integrated over the increasing and decreasing scattering spectral component (integration range $28.5$\,-\,$32.3$\,meV and $32.7$\,-\,$33.5$\,meV respectively), to give a spectral weight function ($SW$) of the high-energy spin-excitation scattering, as plotted for $\Delta S^z$\,=\,$+$\,$1$ in Fig.\,\ref{fig:panel}e. A stepwise ($\tau$\,$<$\,$1.5$\,ps) $SW$ transfer is observed, followed by a $\tau_{\Delta S^z=\pm 1}$\,$>$\,$100$\,ps spin-lattice relaxation component. A zoom-in on the early time-scale spectral dynamics of the $\Delta S^z$\,=\,$+$\,$1$ excitation is plotted in Fig.\,\ref{fig:panel}f. The $SW$ transfer at $t$\,$=$\,$0$\,ps and $t$\,$=$\,$1.5$\,ps amounts to about $15$\% to $30$\% respectively of the $SW$ transfer at late time-delays $t$. This evidences an efficient ultrafast spin disordering mechanism, in agreement with the observations described in Ref.\,\citenum{langner2017}. Similar dynamics is observed for the $\Delta S^z$\,=\,$-$\,$1$ excitation. Comparison of the $\Delta S^z$\,=\,$+$\,$1$ spin excitation peak shift at late delay times with steady-state data allows to calculate a heating of $\Delta T$\,$\approx$\,$7$\,K when quasi-equilibrium is established, in good agreement with the temperature increase estimated from the pulse power, absorption coefficient,\cite{versteeg2016} and low temperature heat capacity of \COSO.\cite{longwavelength,prasai2017} 

The disordering of long-range and internal spin cluster order occurs through different spin-lattice relaxation channels.\cite{kittel1958,kimel2002,kirilyuk2010} 
After the optical excitation the electronically excited system dissipates energy by emission of optical phonons, which subsequently decay into acoustic phonons. \cite{klemens1966,ogasawara2005,perfetti2007} The rapid magnetic spectral weight transfer on the shortest timescale is evidence of an additional decay mechanism for the optical phonons into high-energy spin cluster excitations, which leads to an ultrafast reduction of long-range and internal spin cluster order. Such efficient phonon-magnon decay is enabled by the energy-momentum-dispersion overlap of the optical phonons and the high-energy spin cluster excitations.\cite{kittel1958,portnichenko2016}

\begin{figure}[h!]
\center
\includegraphics[scale=1]{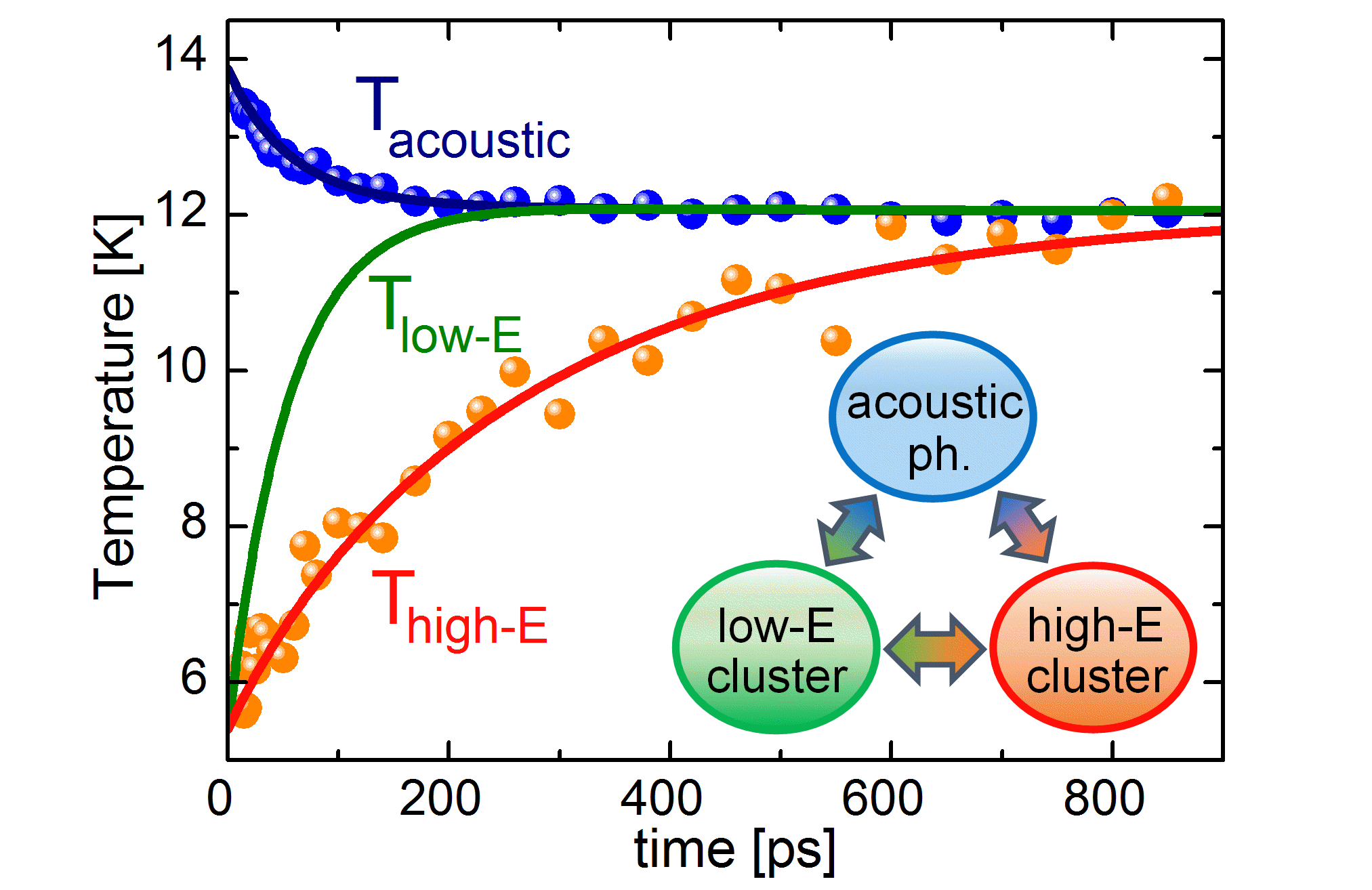}
\caption{Acoustic phonon (ph.), low-energy (low-E), and high-energy (high-E) spin excitation equilibration dynamics after $t$\,$>$\,$6$\,ps. Effective temperatures are indicated with T$_{\rm acoustic}$ for the acoustic phonons, and T$_{\rm low-E}$ and T$_{\rm high-E}$ for the low-energy and high-energy spin cluster excitations.
}
\label{fig:3tm}
\end{figure}

The separation between long-range and internal spin cluster order dynamics becomes apparent over longer timescales. The long timescale spin-lattice equilibration is microscopically dictated by the coupling between acoustic phonons, and low- and high-energy spin cluster excitations. \cite{kimel2002} The low-energy cluster excitation thermalization dynamics can be obtained by fitting a phenomenological three-temperature model \cite{kirilyuk2010} to the phonon and high-energy spin excitation transients, as shown in Fig.\,\ref{fig:3tm}. The change in the acoustic phonon temperature is proportional to the change in the phonon Raman scattering intensity. The change in the high-energy spin excitation temperature is proportional to the observed spin excitation spectral weight transfer. The solid lines are the solution to the three-temperature model. \cite{suppl1,prasai2017} A long-range disordering time of $\tau_{\rm LRO}\sim$\,$55$\,ps is found from the low-energy spin excitation thermalization. The high-energy spin excitations form a dual probe of long-range and internal spin cluster order. From the high-energy spin excitation thermalization we infer an internal cluster disordering time of $\tau_{\rm cluster}\sim$\,$400$\,ps. Similar demagnetization timescales were reported in Ref.\,\citenum{langner2017}.

The separation into double magnetic order parameter dynamics is understood from vastly different phonon-magnon interactions. Acoustic phonons couple strongly to the low-energy spin cluster excitations,\cite{nomura2018phonon} but only weakly to the high-energy spin cluster excitations.\cite{kittel1958} Phonon decay into low-energy spin excitations of the long-range ordered state describes the conventional demagnetization pathway of insulating magnetic materials. \cite{kimel2002,kirilyuk2010} The disordering of internal spin cluster order however has to occur through upconversion of acoustic phonons and/or low-energy spin excitations into high-energy spin cluster excitations, forming a scattering bottleneck in the equilibration dynamics.

\begin{figure}[h!]
\center
\includegraphics[width=3.375in]{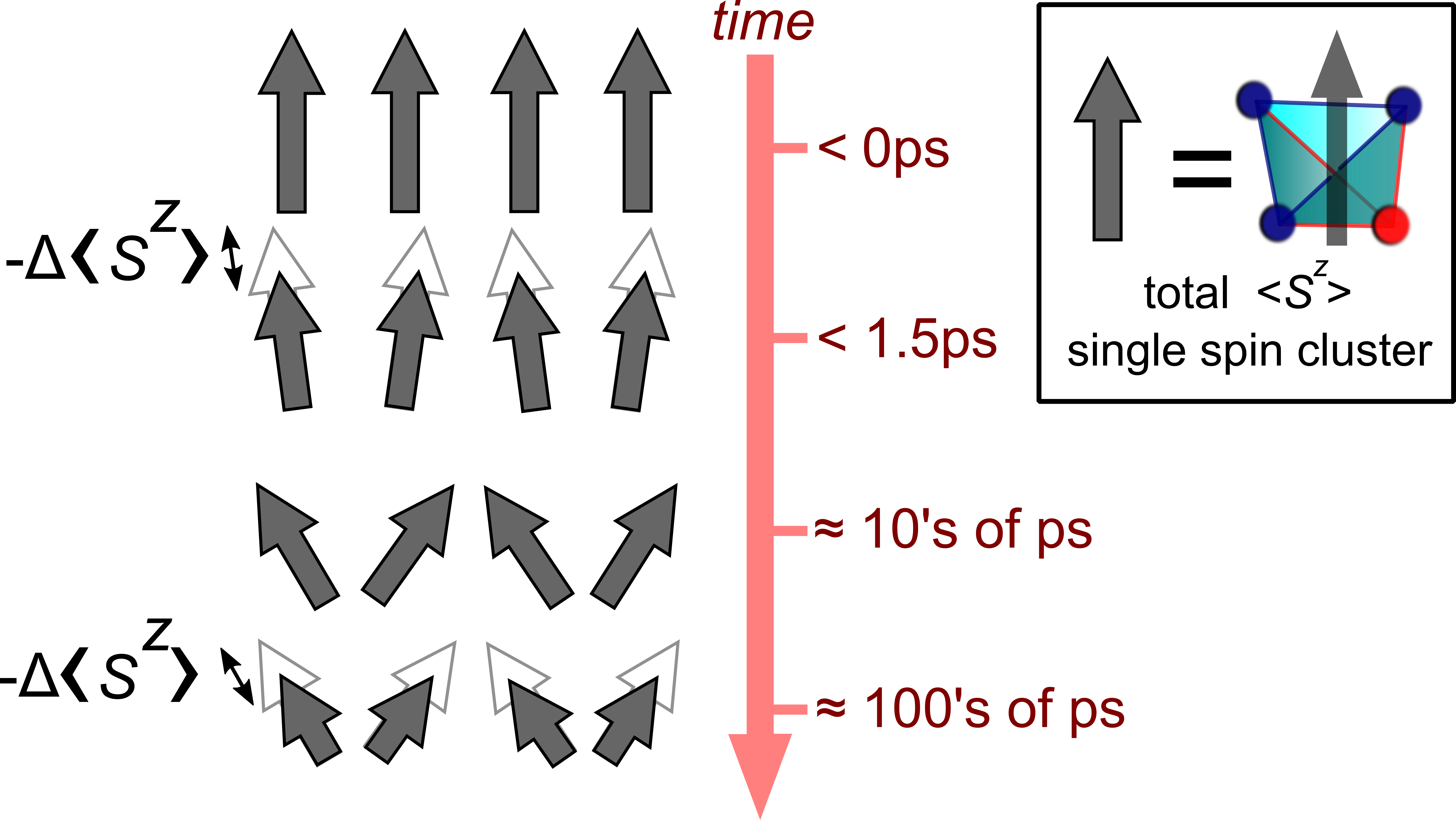}
\caption{Summary of the photoinduced multiple ps-decade spin disordering dynamics. The total spin length $\langle S^z\rangle$ of a single spin cluster is indicated with a grey arrow. The disordering of internal spin cluster order is depicted as a $-\Delta \langle S^z\rangle$ spin length reduction. Long-range disordering is depicted by a reorientation of the cluster spins.}
\label{fig:spinlengthreduction}
\end{figure}

The multiple ps-decade long-range and internal spin cluster order parameter dynamics
is summarized in Fig.\,\ref{fig:spinlengthreduction}. An initial ultrafast ($\tau$\,$<$\,$1.5$\,ps) reduction of long-range and internal spin cluster order results from the decay of optical phonons into high-energy spin cluster excitations. This is depicted by a disordering of the cluster spin alignment and a $-\Delta \langle S^z\rangle$ decrease in spin length for the clusters. On the $10$'s of ps timescale the long-range ordering of spin clusters decreases by decay of acoustic phonons into low-energy spin excitations. On the $100$'s of ps timescale the internal spin cluster order decreases through upconversion of acoustic phonons and/or low-energy spin excitations into high-energy spin cluster excitations.

The present results provide a new viewpoint on the photoinduced nonequilibrium dynamics in the cluster Mott insulator \COSO, and highlight the double magnetic order parameter dynamics of cluster magnets.
Addressing nonequilibrium multiple order parameter dynamics is not only important in cluster magnets like \COSO, but also in contemporary problems in the study of quantum materials consisting of long-range ordered \enquote{molecules}, such as unraveling the nature and speed limit of phase transitions in orbital cluster Mott insulators,\cite{dejong2013speed,wall2018ultrafast} and destabilization of spin-dimer competing phases in spin liquid candidate materials.\cite{alpichshev2015}

This project was partially financed by the Deutsche Forschungsgemeinschaft (DFG) through SFB Grossger\"{a}teantrag INST217/782-1 and SFB-1238 (Projects A02 and B05). RBV acknowledges funding through the Bonn-Cologne Graduate School of Physics and Astronomy (BCGS). RBV thanks D.~Inosov (Dresden, DE), S.~Diehl (Cologne, DE), C.~Kollath (Bonn, DE) and F.~Parmigiani (Trieste, IT) for fruitful discussion.


\end{document}